\documentclass[conf, onecolumn]{ao4elt3}
\usepackage{graphicx}  
\usepackage{float}

\usepackage[varg]{txfonts} 
\usepackage[latin1]{inputenc}

\begin{document}
\pagestyle{plain}
\title{SCExAO AS A PRECURSOR TO AN ELT EXOPLANET DIRECT IMAGING INSTRUMENT}
\author{N. Jovanovic\inst{1}\thanks{nem@naoj.org} \and O. Guyon\inst{1,2} \and F. Martinache\inst{1} \and C. Clergeon\inst{1,3} \and G. Singh\inst{1,3} \and S. Vievard\inst{3} \and
T. Kudo\inst{1} \and V. Garrel\inst{1} \and B. Norris\inst{4} \and P. Tuthill\inst{4} \and P. Stewart\inst{4} \and E. Huby\inst{3} \and G. Perrin\inst{3} \and S. Lacour\inst{3}}
\institute{National Astronomical Observatory of Japan, Subaru Telescope, 650 North A'Ohoku Place, Hilo, HI, 96720, U.S.A. \and Steward Observatory, University of Arizona, Tucson, AZ, 85721, U.S.A. \and LESIA, Observatoire de Paris, Meudon, 5 Place Jules Janssen, 92195, France \and Sydney Institute for Astronomy (SIfA), Institute for Photonics and Optical Science (IPOS), School of Physics, University of Sydney, NSW 2006, Australia}
\abstract{
The Subaru Coronagraphic Extreme AO (SCExAO) instrument consists of a high performance Phase Induced Amplitude Apodisation (PIAA) coronagraph combined with an extreme Adaptive Optics (AO) system operating in the near-infrared (H band). The extreme AO system driven by the $2000$ element deformable mirror will allow for Strehl ratios $>90\%$ to be achieved in the H-band when it goes closed loop. This makes the SCExAO instrument a powerful platform for high contrast imaging down to angular separations of the order of $1\lambda/D$ and an ideal testbed for exploring coronagraphic techniques for ELTs. In this paper we report on the recent progress in regards to the development of the instrument, which includes the addition of a visible bench that makes use of the light at shorter wavelengths not currently utilized by SCExAO and closing the loop on the tip/tilt wavefront sensor. We will also discuss several exciting guest instruments which will expand the capabilities of SCExAO over the next few years; namely CHARIS which is a integral field spectrograph as well as VAMPIRES, a visible aperture masking experiment based on polarimetric analysis of circumstellar disks. In addition we will elucidate the unique role extreme AO systems will play in enabling high precision radial velocity spectroscopy for the detection of small companions.
} 

\maketitle

\section{Introduction}
\label{intro}
The field of high contrast imaging is advancing at a great rate with several
extreme adaptive optics systems set to come online in $2013/2014$ including SCExAO, GPI \cite{macintosh12} and SPHERE \cite{bez08} which will join the already running PALM-$3000$~\cite{op12}. These systems all share a similar underlying architecture: they exploit a high order wavefront sensor (WFS) and a deformable mirror (DM) to correct for atmospheric perturbations enabling high Strehl ratios in the near-infrared (NIR) ($>90\%$), while a coronagraph is used to null the star downstream so that a faint companion can be revealed. Key to achieving contrasts of the order $10^{-6}$ at small angular separations is high level wavefront control and calibration which is the driver behind the need for so-called {\it extreme} adaptive optics. Extreme adaptive optics systems build on conventional single-conjugate adaptive optics systems by correcting for high spatial frequencies as well as the low spatial frequencies. The number of actuators across the pupil of the DM dictates the highest term that can be corrected and hence high element count  (several $1000$ actuator) deformable mirrors are required which have only become available in recent times.

\section{The rebuilt SCExAO instrument}\label{sec:1}
\begin{figure*}[b!]
\centering 
\includegraphics[width=0.90\linewidth]{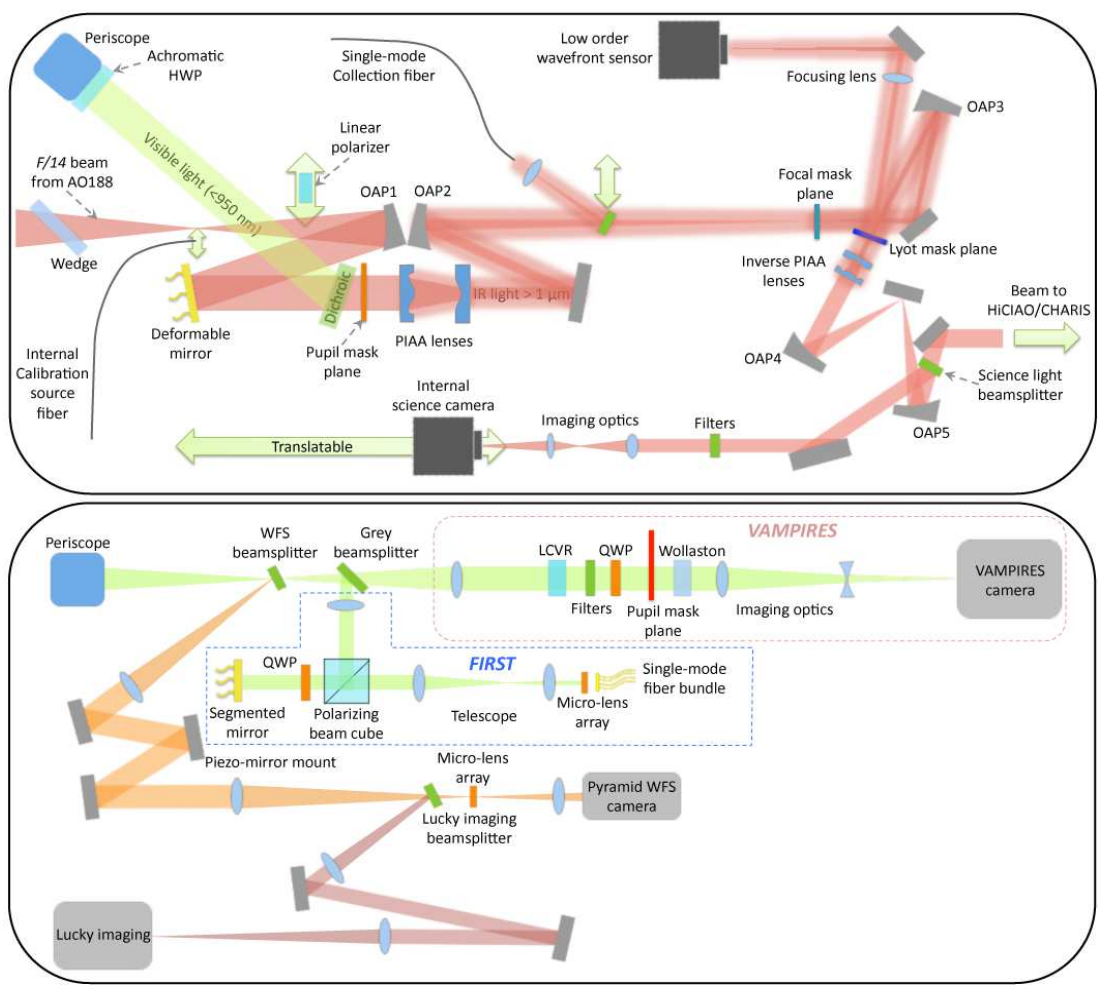}
\caption{\footnotesize The rebuilt SCExAO instrument. Top image: IR optical bench. Dual head green arrows indicate that a given optic can be translated. Bottom image: Visible optical bench which is mounted on top of the IR bench.}
\label{fig:scexao}
\end{figure*}

In this section we outline the SCExAO instrument after the rebuild and upgrade which took place in the summer of $2013$. A detailed schematic of the major components is shown in Fig.~\ref{fig:scexao}. Similar to SPHERE SCExAO aims to exploit unused visible light for scientific exploration. Hence, all light transmitted by the facility adaptive optics system at Subaru, AO$188$, from $600$~nm to $2500$~nm is used. Additional modules that exploit the unused light boost the scientific output of the platform. 
The SCExAO instrument consists of two optical benches stacked on top of one another. The bottom bench is for the IR light modules such as the coronagraphs, deformable mirror and a low order wavefront sensor (LOWFS) (top picture in Fig.~\ref{fig:scexao}) while the top bench is for the visible modules such as the pyramid WFS, VAMPIRES, FIRST and lucky imaging (bottom picture in Figure~\ref{fig:scexao}). The benches are optically connected via a periscope.

Light from AO$188$ is injected into the IR bench of SCExAO with an $f14$ beam where it is first collimated by an off axis parabolic mirror (OAP). All relay optics were changed in the summer of $2013$ to be reflective (gold coated, diamond turned aluminum OAPs) in order to enable achromatic operation. The light is reflected onto the $2000$ element deformable mirror ($2k$ DM) which is situated in the pupil plane. The DM is based on MEMS technology and consists of a thin gold coating on a silicon membrane. The mirror was produced by Boston Micromachines Corporation (BMC) and has $50$ actuators across the $20$~mm diameter and hence $45$ actuators across the $18$~mm pupil. This means that all spatial frequencies out to $22.5\lambda/D$ can be controlled around the Point Spread Function (PSF). On Subaru Telescope ($8$~m primary), this corresponds to a radius for the control region of $0.9"$ in the H-band. The mirror has $5$ dead actuators (can not be modulated), $3$ of which are either outside the pupil or behind the secondary. One is partially occulted by a spider arm and the final one is inside the pupil. For ideal performance these will have to be masked out to prevent leakage of the stellar flux and maximize suppression by the coronagraph. The DM is enclosed in a sealed chamber which is flushed with dry air in order to eliminate high levels of moisture in the vicinity of the mirror in order to prevent corrosion affects. A closed-loop circuit is set up such that if the humidity in the line is greater than $15\%$ the power supply electronics for the DM will be switched off, depowering the DM and prolonging its life.  

The beam reflected off the DM is split by a dichroic into two channels: light shorter than $930$~nm (visible) is reflected up the periscope and onto the top bench while light longer than $950$~nm (NIR) is transmitted.  Once on the top bench the visible light is split between the pyramid wavefront sensor (WFS) and the interferometric imagers (VAMPIRES and FIRST) based on flux and or spectral content. This is adjustable. The light is directed to the pyramid WFS via several mirrors and is focused onto a micro-lens array which acts as the pyramid prism. SCExAO plans to use the pyramid WFS in the non-modulated format which will enable high sensitivity but has a limited range of linearity. As such this sensor relays on a partial correction by AO$188$ to operate effectively. The non-modulated pyramid WFS is currently undergoing laboratory tests and has recently gone closed loop on the first $10$ Zernike modes and $130$ Fourier modes which is encouraging. We hope to have it functioning and fully characterized in the laboratory by the end of $2013$ in time for on-sky commissioning in early $2014$. 

The light transmitted by the dichroic on the IR bench is next transmitted through the various coronagraphs of SCExAO. These are numerous and comprise the Conventional phase induced amplitude apodization (PIAA) coronagraph, the PIAA complex mask coronagraph (PIAACMC), the vector vortex, the $8$-octant phase mask, the $4$-quadrant phase mask and the shaped pupil coronagraphs. Each of the coronagraphs has a corresponding focal plane mask (i.e. the mask that nulls the star) which is designed to diffract the light outside the pupil in the Lyot plane. In the Lyot plane reflective masks direct the diffracted light towards a low order wavefront sensor (LOWFS)~\cite{singh13}. This is used to correct for non-common path and differential chromatic errors between the two benches in regards to tip/tilt and other low order modes. The commands issued by this WFS will be used as an offset to those given by the pyramid WFS. The coronagraphic LOWFS has been shown to reduce the tip/tilt residuals of AO$188$ from $2$~mas to $0.2$~mas on-sky and we believe the new incarnation will be able to reach the same level of performance~\cite{guyon09}. 

The post-coronagraphic light is split at a science beamsplitter wheel by spectral content and/or flux. The light can either be reflected towards an internal science camera or transmitted to a facility science imager such as HiCIAO or CHARIS. The internal science camera is a $320\times256$ pixel, CMOS, InGaAs array which is capable of $348$~Hz frame rates full-frame and has $114~e^{-}$ of read-out noise and low dark current (Axiom Optics - OWL SW$1.7$HS). The frame rate is ideal for coronagraphic imaging and is the key reason why the same detector is used for the LOWFS as well. A set of filters (y, J, H and some narrowband) are located prior to the science camera for tailored science. The camera can be translated between the pupil and focal planes.  

The SCExAO instrument can be aligned/calibrated with the internal calibration source. This consists of a standalone box which hosts a super continuum source (Fianium - Whitelase micro) for broadband characterization, and two fiber coupled laser diodes ($675$~nm and $1550$~nm) for visual alignment. The spectral content and flux level of the broadband source can be controlled and is delivered to the position of the focus of AO$188$ via an endlessly single-mode photonics crystal fiber (NKT photonics - areoGUIDE8). The fiber can be translated into and out of the focus of AO$188$ beam (see Fig.~\ref{fig:scexao}) when internal calibration is preferred or the instrument is not at the telescope. The chosen fiber is ideal as it offers a diffraction-limited point source at all operating wavelengths of the super continuum source and SCExAO ($600-2500$~nm). 

\section{Visible light imagers}
The remainder of the visible light not used by the pyramid WFS is split between a Lucky Fourier imaging module, VAMPIRES and FIRST. The exact splitting can be modified as needed by changing various dichroics. 

\subsubsection{Lucky Fourier Imaging}\label{sec:LFI}
The Lucky imaging camera is used for real-time PSF monitoring. For this a high frame rate ($1$~kHz sub-frame) EMCCD is used (Andor - iXon3 897). In addition to PSF monitoring it can be use for traditional lucky imaging or Fourier lucky imaging~\cite{garrel12}. The technique relies on looking for the strongest Fourier components of each image, and then synthesizing a single image with the extracted Fourier information. In this way diffraction-limited images at $680$~nm of targets like Vega and Betelgeuse  have been synthesized in $2"$ seeing. This is clearly an extremely powerful tool which we will advance by adding multiple spectral channels.

\subsubsection{VAMPIRES}\label{sec:VAMP}
VAMPIRES stands for the Visible Aperture Masking Polarimteric Interferometer for Resolving Exoplanetary Signatures and is installed on the visible bench of SCExAO~\cite{norris12a}. It is designed to share the light not used by the pyramid WFS with the FIRST module. It consists of an aperture masking interferometer which is capable of high contrast imaging, via Fourier imaging techniques, at $<1\lambda/D$~\cite{tuthill00}. This technique has already been used to detect planetary mass companions on these spatial sacles~\cite{kraus12}. Operating in the visible ($600-900$~nm) this interferometer is even more acute to fine spatial structures. VAMPIRES however also has a polarimetric mode. This was motivated by the polarimetric aperture masking work carried out with SAMPOL at the VLT. SAMPOL operates between the y and K bands and was prolific in studying and discovering shells of dust around giant stars that had never been seen before~\cite{norris12b}. Limits where placed on the type and size of the dust grains, but degeneracy between the various models in the NIR meant that this issue could not be clearly resolved. However, the models diverge in the visible and hence data in this region would further constrain dust grain type and size. This was the underlying motivation for the VAMPIRES module. In addition, VAMPIRES has the capabilities to study the dusty disks in exoplantery systems, which is a goal more closely aligned with SCExAO. 

VAMPIRES has 3 levels of polarization calibration. The first is the half wave plate (HWP) mounted to the front of the periscope on the bottom bench. This device is used to rotate the polarization as far upstream as possible to characterize instrument systematics. The fast polarization modulation comes from the Liquid Crystal Variable Retarder (LCVR) which can be modulated at $\sim100$~Hz. This allows for switching of polarizations between images to eliminate quasi-static systematics. Finally the Wollaston prism splits the signal into two distinct inteferograms on the detector with orthogonal polarizations. Imaging is carried out in narrow, $\sim50$~nm bandwidths so as to maintain fringe visibility, but maximize signal-to-noise by using spectral filters.

\subsubsection{FIRST}\label{sec:FIRST}
FIRST is also a visible interferometer based on aperture masking interferometry. However, unlike VAMPIRES which uses a sparse pupil mask, FIRST relies on a $2$D array of single-mode waveguides to collect the light at the pupil plane and reformat it into a non-redundant array for interferometric beam combination. This process referred to as pupil remapping allows for the light across the full pupil to be used enabling higher Fourier coverage. Currently FIRST is using $18$ out of a possible 37 fibers. A micro-lens array is used to couple the light into each optical fiber while a $37$ element segmented mirror upstream, conjugated to the micro-lens array is used for the fine tuning of the tip/tilt for each guide. FIRST has had several successful observing campaigns at the Shane Telescope at Lick Observatory and has now made the move to the Subaru Telescope~\cite{huby12}.

\section{Status}
\begin{figure*}[t!]
\centering 
\includegraphics[width=0.75\linewidth]{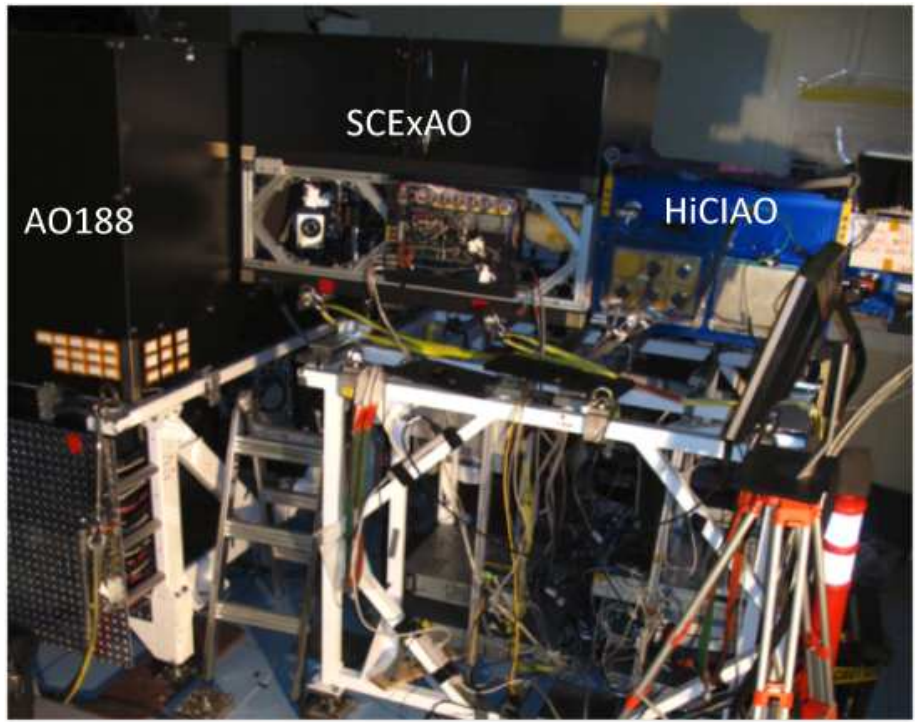}
\caption{\footnotesize The SCExAO instrument at the Nasmyth platform at Subaru Telescope undergoing engineering.}
\label{fig:nas}
\end{figure*}

SCExAO is currently at the end of Phase I engineering and the beginning of Phase II engineering. An image of SCExAO during phase I engineering can be seen in Fig.~\ref{fig:nas}. Phase I milestones include the successful on-sky demonstration of the PIAA coronagraph, the LOWFS and speckle nulling~\cite{martinache12,martinache13b}. Although not discussed here SCExAO offers the ability to suppress quasi-static speckles from one portion of the coronagraphic image by applying sine waves with the correct amplitude, frequency and phase to the DM. This is an active version of imaging techniques such as angular differential imaging (ADI). Since the SCExAO rebuild only the PIAA has been revalidated on-sky. The aim is to revalidate speckle nulling and the LOWFS in upcoming observing runs in December/January of S$13$B.  Phase II engineering goals include closing the loop on the non-modulated pyramid wavefront sensor and achieving a $90\%$ Strehl ratio, as well as validation of VAMPIRES and FIRST.  The non-modulated pyramid WFS is progressing well in the laboratory and we have closed the loop on $\sim130$ Fourier modes thus far. On-sky engineering will be carried out in December/January of S13B. We aim to have it working closed loop consistently by mid to late $2014$. This coincides with a proposed date of moving to open use observing from S$14$B onwards.

VAMPIRES and FIRST had there first on sky night on July the $25^{th}$. They collected much data and are currently analyzing it. They will both receive a single engineering night in early to mid $2014$ and once validated will be made available for open use.

\section{Relevance to ELTs}
SCExAO is a versatile high contrast imaging testbed. Its modular, roll-out commissioning approach allows for new and exciting technologies to be tested. In addition it is one of the only (besides SPHERE) such instruments to exploit light from $600$~nm in the visible to $2500$~nm in the K-band for high contrast imaging. It is also the only platform tailored to high contrast imaging at $<3\lambda/D$ by a host of different technologies. The lessons learnt from this innovative instrument will inform instrument design for future ELTs and indeed alter their science capabilities. A particularly exciting possibility for ELTs is direct imaging and spectroscopic characterization of habitable planets around nearby M-type stars. Around these fainter stars, the planet to star contrast is more favorable than around brighter Sun-like stars, but the small angular separation corresponding to the habitable zone location (closer to the star) will be between $1$ and $3\lambda/D$ on ELT's. SCExAO's focus to develop and validate high contrast imaging techniques for small inner working angles is thus directly aligned with this scientific goal.

\end{document}